\documentclass[%
reprint,
amsmath,amssymb, aps,prl,
]{revtex4-1}

\usepackage{graphicx}
\usepackage{dcolumn}
\usepackage{bm}
\usepackage{hyperref}

\begin{document}

\title{Direct measurement of the hole-nuclear spin interaction in single quantum dots}

\author{E. A. Chekhovich$^{1}$, A. B. Krysa$^2$, M. S. Skolnick$^1$, A. I. Tartakovskii$^1$}

\affiliation{$^{1}$Department of Physics and Astronomy, University
of Sheffield, Sheffield S3 7RH, UK\\
$^{2}$Department of Electronic and Electrical Engineering,
University of Sheffield, Sheffield S1 3JD, UK }

\date{\today}

\begin{abstract}
We use photoluminescence spectroscopy of ''bright'' and ''dark'' exciton states in
single InP/GaInP quantum dots to measure hyperfine interaction of the valence band
hole with nuclear spins polarized along the sample growth axis. The ratio of the
hyperfine constants for the hole ($C$) and electron ($A$) is found to be 
$C/A\approx-0.11$. In InP dots the contribution of spin 1/2 phosphorus nuclei to the hole-nuclear interaction is weak, which enables us to determine experimentally the value of $C$ for spin 9/2 indium nuclei as $C_{In}\approx-5~\mu$eV. This high value of $C$ is in good agreement with recent theoretical predictions and suggests that the hole-nuclear spin interaction has to be taken into account when considering spin qubits based on holes.
\end{abstract}

\maketitle


The spin of an electron confined in a semiconductor quantum dot
has been actively investigated for realization of
solid-state-based quantum bits (qubits). However, the hyperfine
interaction with fluctuating nuclear polarization leads to fast
decoherence of the electron spin on the nanosecond scale
\cite{Khaetskii,Hanson}. In order to circumvent
this problem and realize a solid-state spin-qubit three approaches
are investigated: (i) control of single electron spins in hosts
with zero nuclear spin such as C \cite{Fuchs,Kuemmeth} or Si
\cite{Morton}; (ii) suppression of nuclear spin
fluctuations using elaborate feedback schemes
\cite{Greilich,Reilly,Xu,Latta,Vink} or (iii) use of single holes
instead of electrons, since the contact Fermi coupling with
nuclear spins is zero for the p-type hole wavefunction. Recently slow hole spin relaxation
\cite{Gerardot,Heiss} and long-lived spin coherence \cite{Brunner}
has been demonstrated for InGaAs dots in agreement with the
expected weak hole-nuclear hyperfine interaction. On the other
hand, recent theoretical estimates predict that the
hyperfine interaction of the hole (dipole-dipole in nature) can be as large as 10\% of
that of the electron \cite{Testelin,Fischer}. This interaction has
been used to explain rather fast dephasing of hole spins in an
ensemble of p-doped dots \cite{Eble}, and also the feedback
process leading to suppression of nuclear fluctuations in single
dots in coherent dark-state spectroscopy experiments \cite{Xu}.
However, direct measurement of the hole-nuclear interaction in
quantum dots has not been reported yet.

In this work we use photoluminescence (PL) spectroscopy of single
neutral InP/GaInP quantum dots to directly measure the hole
hyperfine interaction. We measure energy shifts of ''bright'' and
''dark'' excitonic states with all possible electron and heavy
hole spin projections at different magnitudes of optically induced
nuclear spin polarization \cite{Gammon,Chekhovich2}.
This allows accurate measurement of the ratio of the hyperfine
constants of the hole ($C$) and the electron ($A$). We find that
on average $C/A\approx-0.11$. Using the previously measured electron
hyperfine constant $A_{In}\approx47~\mu$eV \cite{Gotschy} and 
taking into account a major contribution of In nuclei into the Overhauser 
shift in InP dots we estimate the heavy hole hyperfine constant $C_{In}\approx-5~\mu$eV.

Our observation of non-zero hole-nuclear spin interaction imply that the 
heavy-light hole mixing, present in most QDs and leading to 
faster hole spin dephasing due to the hyperfine coupling \cite{Eble,Testelin}, 
should be controlled to realize robust QD-based hole-spin qubits. 
We find that when nuclear spins are polarized, holes can
experience effective nuclear magnetic fields on the order of
100~mT. Much weaker magnetic fields have been shown recently to
result in significant enhancement of hole spin coherence \cite{Eble}, 
implying that nuclear spin effects have to be taken into account 
when interpreting experiments on hole spin control.

Our experiments were performed on an undoped InP/GaInP QD sample
without electric gates. PL of neutral InP QDs was measured at
$T=4.2$~K, in external magnetic field $B_z$ up to 8~T normal to
the sample surface. QD PL at $\sim$1.84~eV was excited with a laser at $E_{exc}$=1.88~eV
below the GaInP barrier band-gap and analyzed with a 1~m double spectrometer and a CCD.

\begin{figure}
\includegraphics{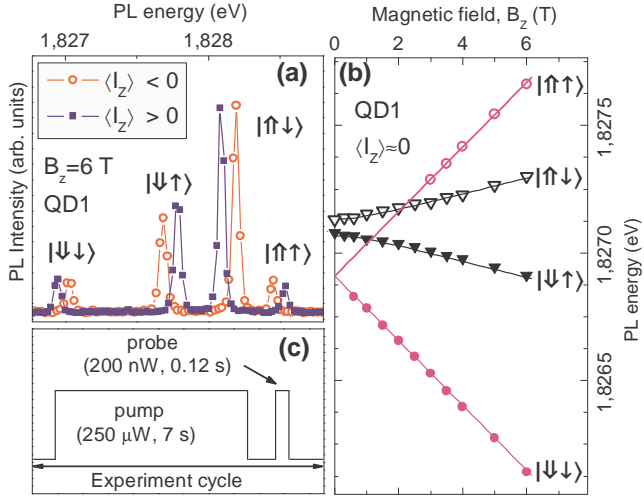}
\caption{\label{fig:Spec} (a) Exciton PL spectra in a neutral
quantum dot at $B_z$=6.0~T. Heavy holes $\Uparrow$($\Downarrow$)
and electrons $\uparrow$($\downarrow$) with spin parallel
(antiparallel) to external field form optically allowed
($\left|\Uparrow\downarrow\right>$,
$\left|\Downarrow\uparrow\right>$) and ''dark''
($\left|\Uparrow\uparrow\right>$,
$\left|\Downarrow\downarrow\right>$) excitons. The presented
spectra correspond to different magnitudes of nuclear spin
polarization $\langle I_z\rangle<0$ ($\bigcirc$) and $\langle
I_z\rangle>0$ ($\blacksquare$). (b) Magnetic field dependence of
exciton PL energies (symbols) at zero nuclear polarization
$\langle I_z\rangle\approx0$. Diamagnetic shift $\kappa=5.8~\mu
eV/T^2$ is subtracted for clarity. Lines show fitting. (c)
Time diagram of the pump-probe experiment cycle. Nuclear spin
$\langle I_z\rangle$ is initialized by a high power pump laser pulse, 
while a low power probe pulse is used to measure PL of both
''bright'' and ''dark'' excitons [as in Fig. (a)].}
\end{figure}

In a neutral dot electrons $\uparrow$($\downarrow$) with spin
$s_z^e=\pm1/2$ and heavy holes $\Uparrow$($\Downarrow$) with
momentum $j_z^h=\pm3/2$ parallel (antiparallel) to the growth axis
$Oz$ can form either optically-forbidden (''dark'') excitons
$\left|\Uparrow\uparrow\right>$
($\left|\Downarrow\downarrow\right>$) with spin projection
$J_z=+2(-2)$, or ''bright'' excitons
$\left|\Uparrow\downarrow\right>$
($\left|\Downarrow\uparrow\right>$) with $J_z=+1(-1)$ optically
allowed in $\sigma^+$($\sigma^-$) polarization.
QD axis misorientation or symmetry reduction leads to weak mixing of ''bright'' and ''dark''
states: as a result the latter are observed in PL
\cite{Bayer,Chekhovich1}. This is demonstrated in Fig. \ref{fig:Spec}
(a) where PL spectra of QD1 measured at low excitation power
$P_{exc}=200$~nW in magnetic field $B_z=6$~T are shown for
different magnitudes of nuclear spin polarization $\langle
I_z\rangle$ (explained below). The dependence of PL energies of
all 4 exciton states measured at different fields $B_z$ is shown
with symbols in Fig. \ref{fig:Spec}(b), their fitting shown with
lines allows to determine electron and hole g-factors:
$g_z^e=1.65$, $g_z^h=2.7$ respectively in QD1 (see appendix in
Ref. \cite{Chekhovich1} for more details on QD characterization).

Non-zero average nuclear spin polarization $\langle I_z\rangle$
along $Oz$ axis acts as an additional magnetic field on the
electron and hole spins. Following Ref.
\cite{Testelin} it is convenient to introduce hole pseudospin
$S_z^h=\pm1/2$ corresponding to the $\Uparrow$($\Downarrow$) heavy
hole state. Coupling of the electron to the nuclei is described by
the hyperfine constant $A$, whereas for the heavy hole the
dipole-dipole interaction with nuclei
\cite{Testelin,Fischer} is described using constant $C$ expressed
in terms of the normalized heavy-hole hyperfine constant $\gamma$
as $C=\gamma A$. The expression for the exciton energy taking into
account the shift due to non-zero average nuclear spin
polarization  can be written as:
\begin{eqnarray}
E[S_z^h,s_z^e]=E^{QD}+E^0[S_z^h,s_z^e]+(s_z^e+\gamma
S_z^h)A\langle I_z\rangle, \label{eq:Ener}
\end{eqnarray}
where the quantum dot band-gap $E^{QD}$ and shift
$E^0[S_z^h,s_z^e]$ determined by the Zeeman and exchange energy
\cite{Bayer} do not depend on nuclear polarization. We
note that Eq. \ref{eq:Ener} is strictly valid only for ''pure''
electron and heavy hole spin states with possible deviations 
arising mainly from the heavy-light hole mixing and
leading to renormalization of $\gamma$ (to be discussed in detail
below). For description of the experimental results we will use
parameter $\gamma^*$ in order to distinguish the hyperfine
constant observed experimentally from the ''pure'' heavy-hole
hyperfine constant $\gamma$.

Since mixing of ''dark'' and ''bright'' excitonic states is weak,
the oscillator strength of the ''dark'' states is small, leading
to their saturation at high powers. As a result, all four exciton
states can be observed in PL only at low excitation power
$P_{exc}\lesssim200$~nW. However, at this low power, optically
induced nuclear spin polarization is small and weakly depends on
polarization of photoexcitation \cite{Chekhovich2}, and thus the
shifts of the hole spin states due to the interaction with nuclei
cannot be measured accurately. In order to avoid this problem, we
use the pump-probe technique \cite{Chekhovich3} with the experiment
cycle shown in Fig. \ref{fig:Spec} (c). Nuclear spin polarization
is prepared with a long ($t_{pump}=7$~s) high power
$P_{exc}=250$~uW pump pulse. After that, the sample is excited
with a low power $P_{exc}=200$~nW probe pulse, during which the PL
spectrum is measured. The duration of this pulse is short enough
($t_{pump}=0.12$~s) to avoid the effect of excitation on nuclear
polarization. The whole cycle is repeated several times
to achieve required signal to noise ratio in PL spectra.

The direct and simultaneous measurement of the hole and electron energy shifts due to the
hyperfine interaction is carried out by detecting the probe spectra recorded
at different magnitudes of $\langle I_z\rangle$ prepared by the pump.
For this, the linearly polarized pump laser first passes through a half-wave plate
followed by a quarter-wave plate.  In order to change $\langle I_z\rangle$, the
half-wave plate is rotated to a new angle $\theta$, leading to a change in the polarization
of the pump, in turn producing a change in spin polarization of the photo-excited electrons in the dot.
For each $\theta$, $\langle I_z\rangle$ reaches the steady-state value proportional to the electron spin polarization. As a result $\langle I_z\rangle$ changes periodically as a function of  $\theta$ \cite{Eble,Skiba}. This is demonstrated in Fig. \ref{fig:Spec} (a)
where the probe spectra measured for $\sigma^+$ ($\langle
I_z\rangle<0$) and $\sigma^-$ ($\langle I_z\rangle>0$) polarized
pump are shown: as expected when $\langle I_z\rangle$ changes, the
exciton states with electron spin $\uparrow$ and $\downarrow$ shift in opposite
directions.

\begin{figure}
\includegraphics{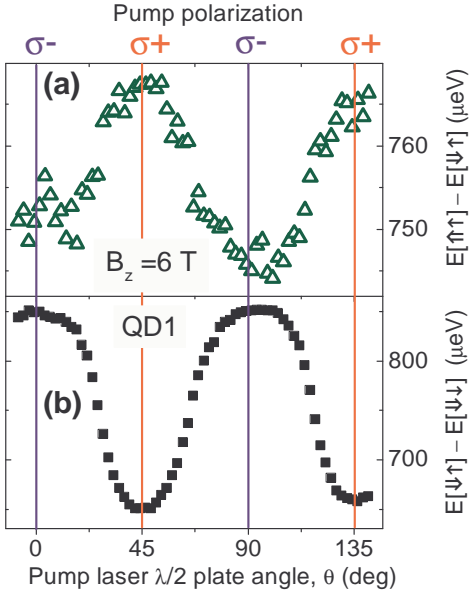}
\caption{\label{fig:Plate} Measurement of the electron- and
hole-nuclear interaction in a neutral dot QD1 at $B_z=6~T$. The
angle $\theta$ of the $\lambda/2$-plate is varied to change
polarization of the pump laser resulting in a change of nuclear
spin polarization $\langle I_z\rangle$. Variation of the splitting between
$\left|\Downarrow\uparrow\right>$ and
$\left|\Downarrow\downarrow\right>$ excitons with $\theta$ [shown in (b)] is
a result of the electron hyperfine interaction and reflects variation of
$\langle I_z\rangle$ induced by the pump. A smaller change of the
splitting between $\left|\Uparrow\uparrow\right>$ and
$\left|\Downarrow\uparrow\right>$ excitons shown in (a) corresponds to
variation of the hole spin splitting and is an evidence for nonzero
hole-nuclear spin interaction.}
\end{figure}

As follows from Eq. \ref{eq:Ener} the energy splitting between
$\left|\Downarrow\uparrow\right>$ and
$\left|\Downarrow\downarrow\right>$ excitons $\Delta
E[\Downarrow\uparrow,\Downarrow\downarrow]=E[\Downarrow\uparrow]-E[\Downarrow\downarrow]\propto
A\langle I_z\rangle$ is determined only by the electron-nuclear
spin interaction, whereas the splitting between
$\left|\Uparrow\uparrow\right>$ and
$\left|\Downarrow\uparrow\right>$ states $\Delta
E[\Uparrow\uparrow,\Downarrow\uparrow]=E[\Uparrow\uparrow]-E[\Downarrow\uparrow]\propto
\gamma A\langle I_z\rangle=C\langle I_z\rangle$ is only due to the
hole-nuclear spin interaction. The dependence of these two
splittings on the angle of the half-wave plate $\theta$ (and
consequently on the value of $\langle I_z\rangle$ induced by the
pump) is shown in Figs. \ref{fig:Plate} (a,b). It can be seen that
the electron spin splitting (Fig. \ref{fig:Spec} (b)) gradually
changes by almost 200~$\mu$eV when the pump polarization is varied
from $\sigma^+$ to $\sigma^-$. At the same time a much weaker
change of the hole spin splitting in antiphase with the electron
spin splitting can be seen in Fig. \ref{fig:Spec} (a) providing a
direct evidence for nonzero hole hyperfine interaction.

In order to obtain a quantitative measure of the hole-nuclear
interaction we fit the experimental results using Eq.
\ref{eq:Ener}. In the fitting we use common values of $\gamma^*$
and $E^0[S_z^h,s_z^e]$, while $E^{QD}$ and $A\langle I_z\rangle$
are varied independently for each position $\theta$ of the
half-wave plate (variation of $E^{QD}$ takes into account spectral
diffusion observed in PL). From this fitting we obtain
$\gamma^*=-0.085\pm0.015$ for QD1.

\begin{figure}
\includegraphics{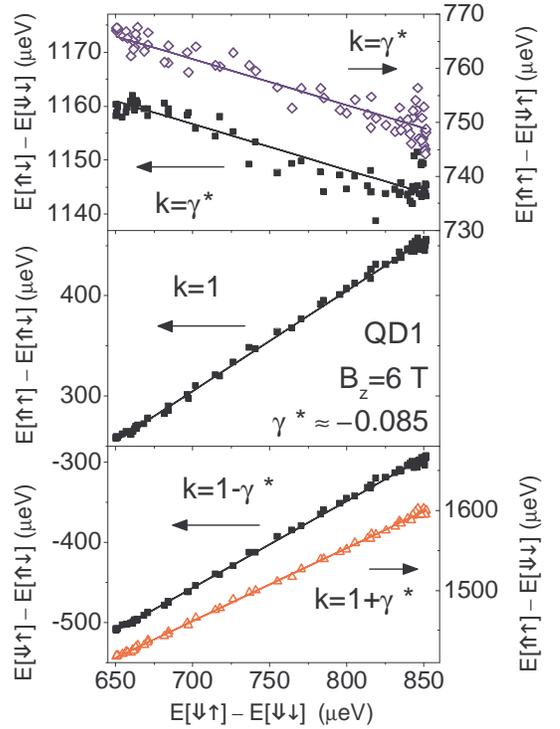}
\caption{\label{fig:Split} Comparison of the experimental results for QD1
with the model of the electron/hole-nuclear spin interaction.
Symbols show experimental dependences of splittings between
different pairs of exciton states on the splitting of
$\left|\Downarrow\uparrow\right>$ and
$\left|\Downarrow\downarrow\right>$ states (i.e. the electron spin
splitting). Straight lines with corresponding coefficients $k$
from Eq. \ref{eq:Split} show fitting with $\gamma^*\approx-0.085$.}
\end{figure}

For direct comparison of experiment with the model we present 
the data in a slightly different way. 
We first note that according to Eq. \ref{eq:Ener} the energy
splitting of any two states is a linear function of the splitting
of any other two states. Choosing $\Delta
E[\Downarrow\uparrow,\Downarrow\downarrow]$ as reference we can
write for all other splittings:
\begin{eqnarray}
\Delta
E[\Downarrow\uparrow,\Uparrow\downarrow]\propto(1-\gamma)\Delta
E[\Downarrow\uparrow,\Downarrow\downarrow]\nonumber\\
\Delta
E[\Uparrow\uparrow,\Downarrow\downarrow]\propto(1+\gamma)\Delta
E[\Downarrow\uparrow,\Downarrow\downarrow]\nonumber\\
\Delta E[\Uparrow\uparrow,\Uparrow\downarrow]\propto\Delta
E[\Downarrow\uparrow,\Downarrow\downarrow]\nonumber\\
\Delta E[\Uparrow\uparrow,\Downarrow\uparrow]\propto\gamma\Delta
E[\Downarrow\uparrow,\Downarrow\downarrow]\nonumber\\
\Delta
E[\Uparrow\downarrow,\Downarrow\downarrow]\propto\gamma\Delta
E[\Downarrow\uparrow,\Downarrow\downarrow]. \label{eq:Split}
\end{eqnarray}
Experimental dependences of these splittings on $\Delta
E[\Downarrow\uparrow,\Downarrow\downarrow]$ are shown in Fig.
\ref{fig:Split} with symbols. Solid lines show linear fitting with
coefficients $k$ determined by Eq. \ref{eq:Split}.
As seen the model involving only one parameter $\gamma^*$
describing the hole-nuclear spin interaction gives a good agreement with
the experiment: the deviation is within $\approx\pm5$~$\mu$eV
mainly determined by the accuracy of PL energy measurement.

\begin{table}[b]
\caption{\label{tab:Dots}%
Experimentally measured hole hyperfine constants  $\gamma^*$, and
circular polarization degrees $\rho_c$ of bright exciton PL for
different neutral QDs at $B_z=6~$T.}
\begin{ruledtabular}
\begin{tabular}{cccc}
\textrm{QD}& \textrm{$\gamma^*$}& \textrm{$\rho_c$}\\
\colrule
QD1 & -0.085$\pm$0.015 & 0.82 \\
QD2 & -0.110$\pm$0.016 & 0.85 \\
QD3 & -0.106$\pm$0.017 & 0.84 \\
QD4 & -0.117$\pm$0.033 & 0.88 \\
QD5 & -0.111$\pm$0.026 & 0.96 \\
QD5 & -0.117$\pm$0.020 & 0.92 \\
\end{tabular}
\end{ruledtabular}
\end{table}

We have performed similar experiments on another 5 neutral dots
from the same sample. 90\% confidence probability estimates
$\gamma^*$ obtained from the fitting using Eq. \ref{eq:Ener} are
given in Table \ref{tab:Dots}. As seen values of $\gamma^*$ coincide 
within the experimental error for all dots, and the average value is
$\bar{\gamma^*}\approx-0.105\pm0.008$.

We will now discuss possible deviations from the model describing pure electron 
and heavy hole states (Eq. \ref{eq:Ener}) and their consequences for
the interpretation of the results presented above.

(i) Bright excitons exhibit fine structure splitting (FSS)
$\delta_b$ at $B_z=0$ and have zero electron spin projections
along $Oz$ axis \cite{Bayer}. Magnetic field $B_z$ partly restores
electron spin projections, at high field
($\delta_b^2/(\mu_Bg_z^eB_z)^2\ll1$) they become
$s_z^e\approx\pm1/2[1-(1/2)\delta_b^2/(\mu_Bg_z^eB_z)^2]$ for
$\left|\Uparrow\downarrow\right>$ and
$\left|\Downarrow\uparrow\right>$ bright excitons. For dark
excitons FSS is much smaller and so $s_z^e\approx\pm1/2$. This
difference will result in violation of the model described by Eq.
\ref{eq:Ener}, in particular the proportionality coefficients in
Eq. \ref{eq:Split} will deviate by
$\approx(1/4)\delta_b^2/(\mu_Bg_z^eB_z)^2$. However, at high
magnetic field $B_z=6$~T the largest correction for the studied
dots (for QD1) is $\approx6\times10^{-3}$. This is smaller than
the uncertainty in  measurements of $\gamma$ and thus can be
neglected.

(ii) Another source of the electron spin projection uncertainty is
mixing of the dark and bright states which we use in this work to
detect the dark excitons. The magnitude of this
mixing can be estimated from the ratio of the maximum PL intensities
of dark and bright states: the maximum intensity is proportional
to the oscillator strengths which for dark states is determined by the
admixture of the bright states \cite{Chekhovich2}. For all dots
this mixing is $<0.01$, negligible compared with our accuracy in determining $\gamma$.

(iii) Finally mixing of heavy holes with $j_z^h=\pm3/2$ and light
holes with $j_z^h=\pm1/2$ must be taken into account. In the
simplest case it leads to the hole spin states of the form
$\left|j_z^h=\pm 3/2\right>+\beta\left|j_z^h=\mp 1/2\right>$ with
$|\beta|\ll1$ \cite{Leger,Krizhanovskii}. It has been
shown, that the hyperfine constant for the light hole interaction
with nuclear spins polarized along $Oz$ is 3 times smaller than
that for the heavy hole \cite{Testelin}. Thus in the case of mixed
hole states the hole hyperfine constant will read as
$\gamma^*=\frac{C}{A}\frac{1-\beta^2/3}{1+\beta^2}=\gamma\frac{1-\beta^2/3}{1+\beta^2}$.
The mixing parameter $\beta$ can be estimated from the circular
polarization degree of PL resulting from recombination of
$\left|j_z^h=\pm 3/2\right>+\beta\left|j_z^h=\mp 1/2\right>$ hole:
$\rho_c=(I^{\sigma^\pm}-I^{\sigma^\mp})/(I^{\sigma^+}+I^{\sigma^-})$,
where $I^{\sigma^{\pm}}$ is PL intensity in $\sigma^{\pm}$
polarizations. In terms of $\beta$,
$\rho_c=(1-\beta^2/3)/(1+\beta^2/3)$ with $\rho_c=1$ for pure
heavy holes \cite{Leger}. Thus the corrected value of $\gamma$
for heavy hole states is expressed as
\begin{eqnarray}
\gamma=\gamma^*(2-\rho_c)/\rho_c, \label{eq:Correction}
\end{eqnarray}
where $\gamma^*<\gamma$ is a value measured experimentally.
$\rho_c$ measured for studied quantum dots at $B_z=$6~T (averaged
for $\left|\Uparrow\downarrow\right>$ and
$\left|\Downarrow\uparrow\right>$ bright excitons)  is shown in
Table \ref{tab:Dots}. Observation of $\rho_c<1$ can be also due to
imperfect shapes of sub-wavelength apertures used to select single
QDs, or imperfections of polarization optics. Taking this into
account and using Eq. \ref{eq:Correction} for each dot we find
that the pure heavy hole hyperfine interaction $\gamma>-0.145$
with 90\% confidence probability. This estimate does not differ
significantly from the average value
$\bar{\gamma^*}\approx-0.105\pm0.008$ for the dots that we have
studied. We thus conclude that the effect of heavy-light hole
mixing is not very strong in the studied structures and
consequently we can use $\bar{\gamma^*}$ as an estimate of
hyperfine interaction for pure heavy holes
$\gamma\approx\bar{\gamma^*}$.

$\gamma$ is an average for interaction with P and In nuclei.
However, contribution of the spin 1/2 P nuclei into the total
Overhauser shift is less then 10\% \cite{Gotschy} as the In nuclei
possess spin 9/2. Since we observe nuclear polarization degree up
to 50\%, contribution of the In nuclei is dominant, and as a
result the estimated value of $\gamma$ corresponds mainly to the
hyperfine interaction with In. Using the value of the electron
hyperfine constant in InP $A_{In}=47$~$\mu$eV  \cite{Gotschy} we
can estimate the heavy hole hyperfine constant
$C_{In}\approx\gamma A_{In}\approx-5$~$\mu$eV. The hyperfine
coupling with In nuclei in different III-V compounds (e.g. InP and
InSb) is similar \cite{Gotschy,Gueron}, and thus this estimate
of $C_{In}$ is applicable in widely studied InGaAs QDs. For the
studied InP dots it is possible to estimate the effective magnetic
field corresponding to fully polarized nuclei: using
experimentally measured g-factors we obtain
$B_{N,max}^e\approx2.4~$T for electrons and
$B_{N,max}^h\approx0.16~$T for heavy holes.

In conclusion, we have employed PL spectroscopy
of neutral excitons in single InP/GaInP quantum dots to
measure the magnitude of the hole-nuclear spin interaction.
On average it is $\bar{\gamma^*}\approx-0.11$ relative to that 
experienced by the electron. It slightly varies from dot to dot, which may be a result of the
varied heavy-light hole mixing and electron-hole overlap.
By measuring the degree of circular polarization of PL, we 
obtain an estimate of the magnitude of heavy-light hole mixing and consequently can 
estimate the hyperfine interaction for the pure heavy hole relative to 
that of the electron as $-0.15\lesssim\gamma\lesssim-0.10$. 

At the final stages of preparation of this manuscript we became aware of differential 
transmission experiments on negatively charged InGaAs dots, where similar magnitudes
of the hole hyperfine constant have been found \cite{Fallahi}.

We thank P. Fallahi for a fruitful discussion. This work has been supported by  the EPSRC Programme Grant EP/G601642/1. 


\end{document}